\documentclass[11pt]{article}
\usepackage{amssymb}
\usepackage{amsmath}
\usepackage{a4wide}
\usepackage{graphicx}

\newtheorem{theorem}{Theorem}

\newtheorem{corollary}{Corollary}

\newenvironment{proof}[1][Proof]{\textbf{#1.} }{\ \rule{0.5em}{0.5em}}
\makeatletter
\def\@removefromreset#1#2{\let\@tempb\@elt
     \def\@tempa#1{@&#1}\expandafter\let\csname @*#1*\endcsname\@tempa
     \def\@elt##1{\expandafter\ifx\csname @*##1*\endcsname\@tempa\else
    \noexpand\@elt{##1}\fi}     \expandafter\edef\csname cl@#2\endcsname{\csname cl@#2\endcsname}     \let\@elt\@tempb
     \expandafter\let\csname @*#1*\endcsname\@undefined}

\@removefromreset{equation}{section}

\@removefromreset{theorem}{section}
\makeatother
\input{tcilatex}

\begin{document}

\title{Threshold bounds for noisy bipartite states}
\author{Elena R. Loubenets\thanks{%
erl@erl.msk.ru}\bigskip  \\
Applied Mathematics Department, Moscow State Institute \\
of Electronics and Mathematics, Moscow 109028, Russia}
\date{}
\maketitle

\begin{abstract}
For a nonseparable bipartite quantum state violating the
Clauser-Horne-Shimony-Holt (CHSH) inequality, we evaluate amounts of noise
breaking the quantum character of its statistical correlations under any
generalized quantum measurements of Alice and Bob. Expressed in terms of the
reduced states, these new threshold bounds can be easily calculated for any
concrete bipartite state. A noisy bipartite state, satisfying the extended
CHSH inequality and the perfect correlation form of the original Bell
inequality for any quantum observables, neither necessarily admits a local
hidden variable model nor exhibits the perfect correlation of outcomes
whenever the same quantum observable is measured on both "sides".
\end{abstract}

\section{Introduction}

The violation of Bell-type inequalities in the quantum case is used in many
quantum information tasks. In reality, one, however, deals with noisy
channels and, for a bipartite quantum state, it is important to estimate
amounts of noise breaking the quantum character of its statistical
correlations under quantum measurements of Alice and Bob.

In the present paper, we analyse this problem based our recent results in
[1, 2] where, in a general setting, we introduced bipartite quantum states, 
\emph{density source operator (DSO) states,} that satisfy the original CHSH
inequality [3] under any generalized quantum measurements of Alice and Bob.
A DSO state with the special dilation property, \emph{a} \emph{Bell class} 
\emph{DSO state},\emph{\ }satisfies\emph{\ }the perfect correlation form of
the original Bell inequality [4] for any three quantum observables\footnote{%
A classical state satisfies a CHSH-form inequality under any Alice and Bob
classical measurements, ideal or randomized, and the perfect correlation
form of the original Bell inequality for any three classical observables
(see appendix of [5] for a general proof of the latter).}.

In section 2, we shortly list the main properties of DSO states specified in
[1, 2] and further prove in a general setting that any DSO state satisfies a
generalized version of the original CHSH inequality [3] - the extended CHSH
inequality, which we introduced in [5].

In section 3, for an arbitrary nonseparable quantum state violating the
original CHSH inequality [3], we introduce the amounts of noise sufficient
for the resulting noisy state to represent a DSO state and a Bell class DSO
state and, therefore, to satisfy the extended CHSH inequality [5] and the
perfect correlation form of the original Bell inequality [4] for any quantum
observables. A noisy bipartite state satisfying the extended CHSH inequality
does not necessarily admit a local hidden variable (LHV)\ model\footnote{%
In the sense formulated by R. Werner in [6].} while a noisy bipartite state
satisfying the perfect correlation form of the original Bell inequality does
not necessarily exhibit the perfect correlation of outcomes if the same
quantum observable is measured on both "sides"\footnote{%
This correlation condition, sufficient for the derivation of the original
Bell inequality, was introduced by J. Bell [4].}.

In section 4, we specify our general bounds for some concrete nonseparable
pure states and, in particular, demonstrate that, satisfying the perfect
correlation form of the original Bell inequality for any three qubit
observables, every separable noisy singlet does not exhibit the perfect
correlation of outcomes whenever the same qubit observable is projectively
measured on both "sides". If, in particular, the same spin observable is
measured on both "sides", the correlation function of a separable noisy
singlet is negative. This explicitly points to the faulty character of the
wide-spread opinion (expressed, for example, in [7]) that the validity of
the perfect correlation form of the original Bell inequality is necessarily
linked with the perfect correlation condition specified by J. Bell [4].

\section{Bipartite quantum states exhibiting classical statistical
correlations}

According to our developments in [1, 2], for any state $\rho $ on a
separable complex Hilbert space $\mathcal{H}_{1}\otimes \mathcal{H}_{2},$
possibly infinite dimensional, there always exist self-adjoint trace class
dilations\footnote{%
The lower indices of $T_{\blacktriangleright },$ $T_{\blacktriangleleft }$
indicate the direction of extension.} $T_{\blacktriangleright }$ on $%
\mathcal{H}_{1}\otimes \mathcal{H}_{2}\otimes \mathcal{H}_{2}$ and $%
T_{\blacktriangleleft }$ on $\mathcal{H}_{1}\otimes \mathcal{H}_{1}\otimes 
\mathcal{H}_{2}$, not necessarily positive, defined by the relations%
\footnote{%
Here, $\mathrm{tr}_{\mathcal{H}_{m}}^{(k)}[\cdot ]$ denotes the partial
trace over the elements of $\mathcal{H}_{m}$ standing in the $k$-th place\
of tensor products.}: $\mathrm{tr}_{\mathcal{H}_{2}}^{(2)}[T_{%
\blacktriangleright }]=\mathrm{tr}_{\mathcal{H}_{2}}^{(3)}[T_{%
\blacktriangleright }]=\rho $ and $\mathrm{tr}_{\mathcal{H}%
_{1}}^{(1)}[T_{\blacktriangleleft }]=\mathrm{tr}_{\mathcal{H}%
_{1}}^{(2)}[T_{\blacktriangleleft }]=\rho .$ This definition implies: $%
\mathrm{tr}[T_{\blacktriangleright }]=\mathrm{tr}[T_{\blacktriangleleft
}]=1. $

We refer to any of these dilations as a \emph{source operator} for a
bipartite state $\rho .$ Since any positive source operator is a density
operator, we specify it as a \emph{density source operator} \emph{(DSO)}.

If, for a bipartite state $\rho $, there exists a density source operator,
we call this $\rho $ a density source operator state or, \emph{a DSO state},
for short. The set of all DSO states on $\mathcal{H}_{1}\otimes \mathcal{H}%
_{2}$ is convex and includes all separable states and a variety of
nonseparable states.

We say that a state $\rho $ on $\mathcal{H}\otimes \mathcal{H}$ is \emph{a
DSO state of the Bell class} if $\rho $ has a density source operator $T$
with the special dilation property $\mathrm{tr}_{\mathcal{H}}^{(1)}[T]=%
\mathrm{tr}_{\mathcal{H}}^{(2)}[T]=$ $\mathrm{tr}_{\mathcal{H}%
}^{(3)}[T]=\rho .$ A Bell class DSO state may be separable or nonseparable.
For example, as we proved in [1], every Werner state [6] on $\mathbb{C}%
^{d}\otimes \mathbb{C}^{d},$ $d\geq 3,$ separable or nonseparable,
represents a Bell class DSO state. A two-qubit Werner state is a DSO state
of the Bell class if it is separable.

The main properties of DSO states, important for applications, concern the
classical character of their statistical correlations under quantum
measurements of Alice and Bob with two measurement settings on each "side".

Consider a generalized Alice/Bob joint quantum measurement, with real-valued
outcomes $\lambda _{1}\in \Lambda _{1},$ $\lambda _{2}\in \Lambda _{2}$ and
performed upon a state $\rho $ on $\mathcal{H}_{1}\otimes \mathcal{H}_{2}$.
Let this joint measurement be specified by a pair\footnote{%
For concreteness, the first argument in a pair refers to a marginal
measurement (say of Alice) with outcomes $\lambda _{1}$ while the second
argument refers to a Bob marginal measurement, with outcomes $\lambda _{2}$.}
$(a,b)$ of measurement settings and described by positive operator-valued (%
\textit{POV}) measures $M_{1}^{(a)}$ and $M_{2}^{(b)}$ of Alice and Bob,
respectively. The joint probability that outcomes $\lambda _{1}$ and $%
\lambda _{2}$ belong to subsets $B_{1}\subseteq \Lambda _{1}$, $%
B_{2}\subseteq \Lambda _{2}$, respectively, has the form\footnote{%
See, for example, in [5] (section 3).}: $\mathrm{tr}[\rho
(M_{1}^{(a)}(B_{1})\otimes M_{2}^{(b)}(B_{2}))].$ The expectation $\langle
\lambda _{1}\lambda _{2}\rangle _{\rho }^{(a,b)}$ of the product $\lambda
_{1}\lambda _{2}$ of the observed outcomes is given by 
\begin{eqnarray}
\langle \lambda _{1}\lambda _{2}\rangle _{\rho }^{(a,b)} &:&=\int_{\Lambda
_{1}\times \Lambda _{2}}\lambda _{1}\lambda _{2}\mathrm{tr}[\rho
(M_{1}^{(a)}(d\lambda _{1})\otimes M_{2}^{(b)}(d\lambda _{2}))]  \label{1} \\
&=&\mathrm{tr}[\rho (W_{1}^{(a)}\otimes W_{2}^{(b)})],  \notag
\end{eqnarray}%
where $W_{1}^{(a)}:=\int_{\Lambda _{1}}\lambda _{1}M_{1}^{(a)}(d\lambda
_{1}) $ and $W_{2}^{(b)}:=\int_{\Lambda _{2}}\lambda
_{2}M_{2}^{(b)}(d\lambda _{2}) $ are bounded quantum observables on $%
\mathcal{H}_{1}$ and $\mathcal{H}_{2},$ respectively.

As we proved in a general setting in [1, 2]:

\noindent (i) a DSO state $\rho $ on $\mathcal{H}_{1}\otimes \mathcal{H}_{2}$
satisfies the original CHSH inequality [3]: 
\begin{equation}
\left\vert \langle \lambda _{1}\lambda _{2}\rangle _{\rho
}^{(a_{1},b_{1})}+\langle \lambda _{1}\lambda _{2}\rangle _{\rho
}^{(a_{1},b_{2})}+\langle \lambda _{1}\lambda _{2}\rangle _{\rho
}^{(a_{2},b_{1})}-\langle \lambda _{1}\lambda _{2}\rangle _{\rho
}^{(a_{2},b_{2})}\right\vert \leq 2,  \label{2}
\end{equation}%
under any generalized quantum measurements of Alice and Bob with real-valued
outcomes $\left\vert \lambda _{1}\right\vert \leq 1,$ $\left\vert \lambda
_{2}\right\vert \leq 1$ of an arbitrary spectral type;

\noindent (ii) a Bell class DSO state $\rho $ on $\mathcal{H}\otimes 
\mathcal{H}$ satisfies the perfect correlation form of the original Bell
inequality [4]: 
\begin{equation}
\left\vert \langle \lambda _{1}\lambda _{2}\rangle _{\rho
}^{(a,b_{1})}-\langle \lambda _{1}\lambda _{2}\rangle _{\rho }^{(a,b_{2})}%
\text{ }\right\vert \leq 1-\langle \lambda _{1}\lambda _{2}\rangle _{\rho
}^{(b_{1},b_{2})},  \label{3}
\end{equation}%
if, under generalized quantum measurements of Alice and Bob with real-valued
outcomes $\left\vert \lambda _{1}\right\vert \leq 1,$ $\left\vert \lambda
_{2}\right\vert \leq 1$, the marginal POV measures obey the correlation
condition 
\begin{equation}
\int_{\Lambda _{1}}\lambda _{1}M_{1}^{(b_{1})}(d\lambda _{1})=\int_{\Lambda
_{2}}\lambda _{2}M_{2}^{(b_{1})}(d\lambda _{2}).  \label{4}
\end{equation}%
This operator relation does not imply the perfect correlation of outcomes,
specified\footnote{%
For short, we further refer to the perfect correlation of outcomes if the
same observable is measured on both "sides" as \emph{Bell's perfect
correlations}.} by J. Bell in [4], and is always fulfilled in case of Alice
and Bob projective measurements of the same quantum observable on both
"sides".

In view of (\ref{1}), inequality (\ref{3}) and condition (\ref{4}) imply
that a Bell class DSO state $\rho $ on $\mathcal{H}\otimes \mathcal{H}$
satisfies\footnote{%
See [1] (theorem 2).} the perfect correlation form of the original Bell
inequality: 
\begin{eqnarray}
\left\vert \mathrm{tr}[\rho (W_{1}\otimes W_{2})]-\mathrm{tr}[\rho
(W_{1}\otimes \widetilde{W}_{2})]\right\vert &\leq &1-\mathrm{tr}[\rho
(W_{2}\otimes \widetilde{W}_{2})],  \label{5} \\
\left\vert \mathrm{tr}[\rho (W_{1}\otimes W_{2})]-\mathrm{tr}[\rho (%
\widetilde{W}_{1}\otimes W_{2})]\right\vert &\leq &1-\mathrm{tr}[\rho
(W_{1}\otimes \widetilde{W}_{1})],  \notag
\end{eqnarray}%
for any bounded quantum observables $W_{1},$ $\widetilde{W}_{1},$ $W_{2}$, $%
\widetilde{W}_{2}$ on $\mathcal{H}$ with operator norms $\left\Vert \cdot
\right\Vert \leq 1.$ In the right-hand sides of inequalities (\ref{5}),
observables can be interchanged.

A general condition sufficient for an arbitrary DSO state to satisfy (\ref{5}%
) is introduced in [2] (section 3, theorem 4).

In appendix of this paper, extending further our above results [1,2] on DSO\
states, we prove that a DSO state $\rho $ on $\mathcal{H}_{1}\otimes 
\mathcal{H}_{2}$ satisfies a generalized version of (\ref{2}) - the extended
CHSH inequality: 
\begin{equation}
\left\vert \text{ }\gamma _{11}\langle \lambda _{1}\lambda _{2}\rangle
_{\rho }^{(a_{1},b_{1})}+\gamma _{12}\langle \lambda _{1}\lambda _{2}\rangle
_{\rho }^{(a_{1},b_{2})}+\gamma _{21}\langle \lambda _{1}\lambda _{2}\rangle
_{\rho }^{(a_{2},b_{1})}+\gamma _{22}\langle \lambda _{1}\lambda _{2}\rangle
_{\rho }^{(a_{2},b_{2})}\right\vert \leq 2\max_{i,k}|\gamma _{ik}|,
\label{6}
\end{equation}%
which we introduced in [5]. Here, $\gamma _{ik}$ are any real coefficients
obeying either of the relations: 
\begin{equation}
\gamma _{11}\gamma _{12}=-\gamma _{21}\gamma _{22}\text{ \ \ \ or \ \ \ }%
\gamma _{11}\gamma _{21}=-\gamma _{12}\gamma _{22}\text{ \ \ \ or \ \ \ }%
\gamma _{11}\gamma _{22}=-\gamma _{12}\gamma _{21}.  \label{7}
\end{equation}%
Note that the extended CHSH inequality cannot be, in general, derived by
rescaling\footnote{%
Any rescaling of (\ref{2}) results in a version of (\ref{6}) with
coefficients $\gamma _{ik}$ satisfying only the last relation in (\ref{7}).}
of the original CHSH inequality (\ref{2}).

It should be stressed that a nonseparable DSO state does not necessarily
admit an LHV model formulated in [6] while a Bell class DSO state, separable
or nonseparable, does not necessarily exhibit Bell's perfect correlations
(see footnote 8).

\section{Noisy bipartite quantum states}

For an arbitrary state $\rho $ on $\mathbb{C}^{d_{1}}\otimes \mathbb{C}%
^{d_{2}},$ $\forall d_{1},d_{2}\geq 2,$ violating the original CHSH
inequality (\ref{2}), let us now specify the amounts of noise sufficient for
the noisy state $\eta _{\rho }(\beta )=\beta \frac{I_{\mathbb{C}%
^{d_{1}}\otimes \mathbb{C}^{d_{2}}}}{d_{1}d_{2}}+(1-\beta )\rho ,$ $\beta
\in (0,1],$ to satisfy inequalities (\ref{2}), (\ref{3}) and (\ref{6}).

Consider first such Alice and Bob generalized quantum measurements with
real-valued outcomes $\left\vert \lambda _{1}\right\vert \leq 1,$ $%
\left\vert \lambda _{2}\right\vert \leq 1$, where all product averages in
the state $\sigma _{noise}=\frac{I_{\mathbb{C}^{d_{1}}\otimes \mathbb{C}%
^{d_{2}}}}{d_{1}d_{2}}$ are equal to zero: $\langle \lambda _{1}\lambda
_{2}\rangle _{\sigma _{noise}}^{(a_{i},b_{k})}=0,$ $\forall i,k=1,2$. Under
these joint measurements, the noisy state $\eta _{\rho }(\beta )$ satisfies
the original CHSH inequality (\ref{2}) whenever\footnote{%
Here, we take into account that, in the quantum case, the maximal value of
the left-hand side of (\ref{2}) constitutes $2\sqrt{2}$ (the Tsirelson bound
[8]).} $2\sqrt{2}(1-\beta )\leq 2$ $\Leftrightarrow \beta \geq 1-\frac{\sqrt{%
2}}{2}$ and the extended CHSH inequality (\ref{6}) if $4(1-\beta )\leq 2$ $%
\Leftrightarrow \beta \geq \frac{1}{2}.$

If, furthermore, $\langle \lambda _{1}\lambda _{2}\rangle _{\sigma
_{noise}}^{(b_{1},b_{2})}=0,$ then the noisy state $\eta _{\rho }(\beta )$
satisfies\ both forms\footnote{%
The perfect anticorrelation form of the original Bell inequality [4]
corresponds to plus sign in the right-hand sides of (\ref{3}), (\ref{5}).}
of the original Bell inequality (\ref{3}) whenever $2(1-\beta )\leq
1-(1-\beta )$ $\Leftrightarrow \beta \geq \frac{2}{3}$.

However, the above threshold bounds do not need to hold if, under Alice and
Bob joint measurements, at least one of the product averages $\langle
\lambda _{1}\lambda _{2}\rangle _{\sigma _{noise}}^{(a_{i},b_{k})}\neq 0$.
Below, we specify the threshold bounds that are valid under any generalized
quantum measurements of Alice and Bob.

\begin{theorem}
Let a nonseparable state $\rho $ on $\mathbb{C}^{d_{1}}\otimes \mathbb{C}%
^{d_{2}},$ $d_{1},d_{2}\geq 2,$ violate the original CHSH inequality (\ref{2}%
). Denote by\footnote{$||\cdot ||$ denotes the operator norm.} 
\begin{equation}
\gamma _{\rho }:=\min \{d_{1}||\tau _{\rho }^{(1)}||,d_{2}||\tau _{\rho
}^{(2)}||\}\geq 1  \label{8}
\end{equation}%
the parameter characterizing $\rho $ in terms of its reduced states $\tau
_{\rho }^{(1)}:=\mathrm{tr}_{\mathbb{C}^{d_{2}}}[\rho ],$ $\tau _{\rho
}^{(2)}:=\mathrm{tr}_{\mathbb{C}^{d_{1}}}[\rho ]$ on $\mathbb{C}^{d_{1}}$
and $\mathbb{C}^{d_{2}}$, respectively. The noisy state 
\begin{equation}
\eta _{\rho }(\beta )=\beta \frac{I_{\mathbb{C}^{d_{1}}\otimes \mathbb{C}%
^{d_{2}}}}{d_{1}d_{2}}+(1-\beta )\rho ,\text{ \ \ \ }\beta \in \lbrack \beta
_{_{CHSH}}(\rho ),1],\text{ \ \ \ }\beta _{_{CHSH}}(\rho ):=\frac{\gamma
_{\rho }}{1+\gamma _{\rho }},  \label{9}
\end{equation}%
satisfies the extended CHSH inequality (\ref{6}) under any generalized
quantum measurements of Alice and Bob. In (\ref{9}), $\beta _{_{CHSH}}(\rho
)\geq \frac{1}{2},$ $\mathbf{\forall }\rho $.
\end{theorem}

\begin{proof}
Consider the decomposition $\rho =\sum \gamma _{nm,n_{1}m_{1}}|e_{n}\rangle
\langle e_{n_{1}}|\otimes |f_{m}\rangle \langle f_{m_{1}}|$ in orthonormal
bases $\{e_{n}\}$ in $\mathbb{C}^{d_{1}}$ and $\{f_{m}\}$ in $\mathbb{C}%
^{d_{2}}.$ For a noisy state $\eta _{\rho }(\beta ),$ the operator 
\begin{eqnarray}
T_{\blacktriangleright }^{(\beta )}=(1-\beta )\sum \gamma
_{nn_{_{1}},mm_{1}}|e_{n}\rangle \langle e_{n_{_{1}}}|\otimes
\{|f_{m}\rangle \langle f_{m_{_{1}}}|\otimes \xi _{2}+\xi _{2}\otimes
|f_{m}\rangle \langle f_{m_{_{1}}}|\}  \label{10} \\
-(1-\beta )\tau _{\rho }^{(1)}\otimes \xi _{2}\otimes \xi _{2}+\beta \frac{%
I_{\mathbb{C}^{d_{1}}\otimes \mathbb{C}^{d_{2}}\otimes \mathbb{C}^{d_{2}}}}{%
d_{1}d_{2}^{2}}  \notag
\end{eqnarray}
on $\mathbb{C}^{d_{1}}\otimes \mathbb{C}^{d_{2}}\otimes \mathbb{C}^{d_{2}}$
represents a source operator. Here, $\xi _{2}$ is a density operator on $%
\mathbb{C}^{d_{2}}.$

If, in (\ref{10}), the operator $Y^{(\beta )}:=\beta \frac{I_{\mathbb{C}%
^{d_{1}}\otimes \mathbb{C}^{d_{2}}\otimes \mathbb{C}^{d_{2}}}}{d_{1}d_{2}^{2}%
}-$ $(1-\beta )\tau _{\rho }^{(1)}\otimes \xi _{2}\otimes \xi _{2}$ is
nonnegative then the source-operator $T_{\blacktriangleright }^{(\beta )}$
is positive, that is, represents a density source operator. In view of the
relation $-\left\| W\right\| I_{\mathcal{K}}\leq W\leq \left\| W\right\| I_{%
\mathcal{K}}$, valid for any bounded observable $W$ on a Hilbert space $%
\mathcal{K}$, we derive: 
\begin{equation}
Y^{(\beta )}\geq \frac{\beta -d_{1}d_{2}^{2}||\tau _{\rho }^{(1)}||\left\|
\xi _{2}\right\| ^{2}(1-\beta )}{d_{1}d_{2}^{2}}I_{\mathbb{C}^{d_{1}}\otimes 
\mathbb{C}^{d_{2}}\otimes \mathbb{C}^{d_{2}}}.  \label{11}
\end{equation}
Therefore, $T_{\blacktriangleright }^{(\beta )}$ is a density source
operator for any 
\begin{equation}
\beta \geq \frac{d_{1}d_{2}^{2}||\tau _{\rho }^{(1)}||\left\| \xi
_{2}\right\| ^{2}}{1+d_{1}d_{2}^{2}||\tau _{\rho }^{(1)}||\left\| \xi
_{2}\right\| ^{2}}.  \label{12}
\end{equation}
Quite similarly, we construct the source operator $T_{\blacktriangleleft
}^{(\beta )}$ on $\mathbb{C}^{d_{1}}\otimes \mathbb{C}^{d_{1}}\otimes 
\mathbb{C}^{d_{2}}$ and prove that $T_{\blacktriangleleft }^{(\beta )}$ is a
density source operator for any $\beta \geq \frac{d_{1}^{2}d_{2}||\tau
_{\rho }^{(2)}||\left\| \xi _{1}\right\| ^{2}}{1+d_{1}^{2}d_{2}||\tau _{\rho
}^{(2)}||\left\| \xi _{1}\right\| ^{2}},$ where $\xi _{1}$ is a density
operator on $\mathbb{C}^{d_{1}}.$ Taking into account that, for $x\geq 0,$
the function $\frac{x}{1+x}$ is monotone increasing and that, for a density
operator $\xi $ on $\mathbb{C}^{d},$ its operator norm $\frac{1}{d}\leq
\left\| \xi \right\| \leq 1$, we choose in (\ref{11}), (\ref{12}) density
operators $\xi _{1},$ $\xi _{2}$ with $\left\| \xi _{1}\right\| =\frac{1}{%
d_{1}}$, $\left\| \xi _{2}\right\| =\frac{1}{d_{2}}$.

Introducing further parameter (\ref{8}) and noting that $\min \{\frac{%
d_{1}||\tau _{\rho }^{(1)}||}{1+d_{1}||\tau _{\rho }^{(1)}||},\frac{%
d_{2}||\tau _{\rho }^{(2)}||}{1+d_{2}||\tau _{\rho }^{(2)}||}\}=\frac{\gamma
_{\rho }}{1+\gamma _{\rho }},$ we derive that, for any $\beta \geq \frac{%
\gamma _{\rho }}{1+\gamma _{\rho }},$ the noisy state $\eta _{\rho }(\beta )$
is a DSO state and, therefore, satisfies (\ref{6}). Since $d_{1}||\tau
_{\rho }^{(1)}||$ $\geq 1,$ $d_{2}||\tau _{\rho }^{(2)}||$ $\geq 1,$ the
parameter $\gamma _{\rho }\geq 1$ and, hence, $\frac{\gamma _{\rho }}{%
1+\gamma _{\rho }}\geq \frac{1}{2}$.
\end{proof}

\begin{theorem}
Let a state $\rho $ on $\mathbb{C}^{d}\otimes \mathbb{C}^{d},$ $d\geq 2,$
separable or nonseparable, with equal reduced $\tau _{\rho }^{(1)}=\tau
_{\rho }^{(2)}=\tau _{\rho }$, violate the perfect correlation form (\ref{5}%
) of the original Bell inequality. The noisy state 
\begin{equation}
\eta _{\rho }(\beta )=\beta \frac{I_{\mathbb{C}^{d}\otimes \mathbb{C}^{d}}}{%
d^{2}}+(1-\beta )\rho ,\text{ \ \ \ }\beta \in \lbrack \beta _{_{Bell}}(\rho
),1],\text{ \ \ \ }\beta _{_{Bell}}(\rho ):=\frac{2\gamma _{\rho }^{3}}{%
1+2\gamma _{\rho }^{3}},  \label{13}
\end{equation}%
(where $\gamma _{\rho }:=d\left\Vert \tau _{\rho }\right\Vert \geq 1$)
satisfies inequality (\ref{5}) for any three quantum observables on $\mathbb{%
C}^{d}$ and inequality (\ref{3}) under any generalized quantum measurements
of Alice and Bob where the marginal POV measures obey the correlation
condition (\ref{4}). In (\ref{13}), $\beta _{_{Bell}}(\rho )\geq \frac{2}{3}%
, $\ $\forall \rho .$
\end{theorem}

\begin{proof}
Consider the decomposition $\rho =\sum \gamma _{nm,n_{1}m_{1}}|e_{n}\rangle
\langle e_{n_{1}}|\otimes |e_{m}\rangle \langle e_{m_{1}}|$ in an
orthonormal basis $\{e_{n}\}$ in $\mathbb{C}^{d}.$ For a noisy state $\eta
_{\rho }(\beta )$, the operator 
\begin{align}
T^{(\beta )}& =(1-\beta )\sum \gamma _{nn_{1},mm_{1}}\{\text{ }|e_{n}\rangle
\langle e_{n_{1}}|\otimes |e_{m}\rangle \langle e_{m_{1}}|\otimes \tau
_{\rho }+|e_{n}\rangle \langle e_{n_{1}}|\otimes \tau _{\rho }\otimes
|e_{m}\rangle \langle e_{m_{1}}|  \label{14} \\
& +\tau _{\rho }\otimes |e_{n}\rangle \langle e_{n_{1}}|\otimes
|e_{m}\rangle \langle e_{m_{1}}|\}-2(1-\beta )\tau _{\rho }\otimes \tau
_{\rho }\otimes \tau _{\rho }+\beta \frac{I_{\mathbb{C}^{d}\otimes \mathbb{C}%
^{d}\otimes \mathbb{C}^{d}}}{d^{3}}  \notag
\end{align}%
on $\mathbb{C}^{d}\otimes \mathbb{C}^{d}\otimes \mathbb{C}^{d}$ represents a
source operator with the special dilation property 
\begin{equation}
\mathrm{tr}_{\mathbb{C}^{d}}^{(k)}[T^{(\beta )}]=\eta _{\rho }(\beta ),\text{
\ \ \ }k=1,2,3.  \label{15}
\end{equation}%
If, in the left-hand side of (\ref{14}), the operator $\widetilde{Y}^{(\beta
)}:=\beta \frac{I_{\mathbb{C}^{d}\otimes \mathbb{C}^{d}\otimes \mathbb{C}%
^{d}}}{d^{3}}-2(1-\beta )\tau _{\rho }\otimes \tau _{\rho }\otimes \tau
_{\rho }$ is nonnegative then the source-operator $T^{(\beta )}$ is
positive, that is, represents a density source operator. Evaluating $%
\widetilde{Y}^{(\beta )}$ quite similarly as in (\ref{11}), we derive: $%
\widetilde{Y}^{(\beta )}\geq \frac{\beta -2d^{3}||\tau _{\rho
}||^{3}(1-\beta )}{d^{3}}I_{\mathbb{C}^{d}\otimes \mathbb{C}^{d}\otimes 
\mathbb{C}^{d}}$. Therefore, for any $\beta \geq \frac{2\gamma _{\rho }^{3}}{%
1+2\gamma _{\rho }^{3}},$ the source operator $T^{(\beta )},$ with the
special dilation property (\ref{15}), is a density source operator. This
means that the noisy state $\eta _{\rho }(\beta ),$ $\beta \in \lbrack \frac{%
2\gamma _{\rho }^{3}}{1+2\gamma _{\rho }^{3}},1],$ is a Bell class DSO state
and, therefore, satisfies inequality (\ref{5}) for any three quantum
observables on $\mathbb{C}^{d}$ and inequality (\ref{3}) under the condition
(\ref{4}). We have: $\frac{2\gamma _{\rho }^{3}}{1+2\gamma _{\rho }^{3}}\geq 
\frac{\gamma _{\rho }}{1+\gamma _{\rho }}$ and $\frac{2\gamma _{\rho }^{3}}{%
1+2\gamma _{\rho }^{3}}\geq \frac{2}{3}.$\bigskip
\end{proof}

Note that if $\rho $ is an arbitrary state on $\mathcal{H}\otimes \mathcal{H}
$ satisfying the original Bell inequality (\ref{5}) for any bounded quantum
observables on $\mathcal{H},$ then this $\rho $ does not violate the
extended CHSH inequality (\ref{6}) and, in particular, the original CHSH
inequality (\ref{2}). Hence, for a state $\rho $ on $\mathcal{H}\otimes 
\mathcal{H}$, the validity of the original CHSH inequality (\ref{2}) is
necessary for the validity of the original Bell inequality (\ref{5}).
Theorems 1 and 2 imply.

\begin{corollary}
Let a nonseparable state $\rho $ on $\mathbb{C}^{d}\otimes \mathbb{C}^{d},$ $%
d\geq 2,$ with equal reduced $\tau _{\rho }^{(1)}=\tau _{\rho }^{(2)}=\tau
_{\rho }$, violate the original CHSH inequality (\ref{2}). The noisy state $%
\beta \frac{I_{\mathbb{C}^{d}\otimes \mathbb{C}^{d}}}{d^{2}}+(1-\beta )\rho
, $ $\beta \in (0,1],$ does not violate the extended CHSH inequality (\ref{6}%
) if $\beta \geq \beta _{_{CHSH}}(\rho ):=\frac{\gamma _{\rho }}{1+\gamma
_{\rho }}$ and does not violate the perfect correlation form (\ref{5}) of
the Bell inequality whenever $\beta \geq \beta _{_{Bell}}(\rho ):=\frac{%
2\gamma _{\rho }^{3}}{1+2\gamma _{\rho }^{3}}.$ We have: $\beta
_{_{CHSH}}(\rho )\geq \frac{1}{2},$ $\beta _{_{Bell}}(\rho )\geq \frac{2}{3}$
and $\beta _{_{Bell}}(\rho )>\beta _{_{CHSH}}(\rho ).$
\end{corollary}

The threshold bounds specified in theorems 1, 2 and corollary 1 are
sufficient. Therefore, for a nonseparable bipartite state, the threshold
amounts of noise $\beta _{_{CHSH}},$ $\beta _{_{Bell}}$, defined by (\ref{9}%
) and (\ref{13}), are not necessarily least.

\section{Examples}

Consider on\ $\mathbb{C}^{d}\otimes \mathbb{C}^{d},$ $d\geq 2,$ the
nonseparable pure state 
\begin{equation}
\rho _{d,\vartheta }=|\phi _{d,\vartheta }\rangle \langle \phi _{d,\vartheta
}|,\text{ \ \ \ }\phi _{d,\vartheta }=\frac{1}{\sqrt{d}}\sum_{n=1}^{d}\exp
(i\vartheta _{n})e_{n}\otimes e_{n},  \label{16}
\end{equation}
specified by a sequence $\vartheta =\{\vartheta _{n}\}$ of real phases. This
state violates\footnote{%
If all $\vartheta _{n}=0,$ then this state violates (\ref{2}) maximally.}
the original CHSH inequality (\ref{2}) and, therefore, the original Bell
inequality (\ref{5}).

For the nonseparable state $\rho _{d,\vartheta },$ both reduced states are
equal to $\frac{I_{\mathbb{C}^{d}}}{d},$ the parameter $\gamma _{\rho
_{d,\vartheta }}=1$ and, due to corollary 1, the threshold amounts of noise
constitute: $\beta _{_{CHSH}}(\rho _{d,\vartheta })=\frac{1}{2},$ $\beta
_{_{Bell}}(\rho _{d,\vartheta })=\frac{2}{3}.$

In view of the Peres separability criterion [9], for any $\beta \in \lbrack
0,\frac{d}{d+1}),$ the state $\eta _{\rho _{d,\vartheta }}(\beta )=\beta 
\frac{I_{\mathbb{C}^{d}\otimes \mathbb{C}^{d}}}{d^{2}}+(1-\beta )\rho
_{d,\vartheta }$ is nonseparable\footnote{%
The partial transpose of $\eta _{\rho _{d,\vartheta }}(\beta )$ has the
eigenvalue $\frac{\beta (d+1)-d}{d^{2}}$, which is negative for any $\beta <%
\frac{d}{d+1}.$}. Hence, for\textbf{\ }any separable noisy state $\eta
_{\rho _{d,\vartheta }}(\beta ),$ the value of $\beta $ cannot be less than $%
\frac{d}{d+1}.$ Since $\beta _{_{Bell}}(\rho _{d,\vartheta })=\frac{2}{3}$
and $\frac{d}{d+1}\geq \frac{2}{3},$ $\forall d\geq 2,$ we conclude that 
\emph{every separable admixture of }$\rho _{d,\vartheta }$ \emph{to }$\frac{%
I_{\mathbb{C}^{d}\otimes \mathbb{C}^{d}}}{d^{2}}$ \emph{does not violate the
perfect correlation form (\ref{5})\ of the original Bell inequality.}

The latter property holds also for any separable noisy Bell state on $%
\mathbb{C}^{2}\otimes \mathbb{C}^{2}.$ Recall that the Bell states have the
form: 
\begin{equation}
\phi ^{(\pm )}=\frac{1}{\sqrt{2}}(e_{1}\otimes e_{1}\pm e_{2}\otimes e_{2}),%
\text{ \ \ \ }\psi ^{(\pm )}=\frac{1}{\sqrt{2}}(e_{1}\otimes e_{2}\pm
e_{2}\otimes e_{1}),  \label{17}
\end{equation}%
and that a noisy Bell state is separable iff $\beta \geq \frac{2}{3}.$ For
any of the Bell states (\ref{17}), the reduced states are given by $\frac{I_{%
\mathbb{C}^{2}}}{2}$, parameter (\ref{8}) is equal to one and, therefore,
due to corollary 1, the threshold amounts of noise $\beta _{_{CHSH}}=\frac{1%
}{2},$ $\beta _{_{Bell}}=\frac{2}{3}$.

Thus, \emph{every separable admixture of a Bell state to} $\frac{I_{\mathbb{C%
}^{2}\otimes \mathbb{C}^{2}}}{4}$ $\emph{does}$ $\emph{not}$ $\emph{violate}$
$\emph{the}$ $\emph{perfect}$ $\emph{correlation}$ $\emph{form}$ \emph{(\ref%
{5})} \emph{of the original Bell inequality.}

Note that the noisy singlet $\eta _{\psi ^{(-)}}(\frac{1}{2})=\frac{I_{%
\mathbb{C}^{2}\otimes \mathbb{C}^{2}}}{8}+\frac{|\psi ^{(-)}\rangle \langle
\psi ^{(-)}|}{2},$ corresponding to the threshold value $\beta _{_{CHSH}}=%
\frac{1}{2},$ represents the two-qubit nonseparable Werner state [6],
specified in [6] by the parameter $\Phi =-\frac{1}{4}$.

Let us now explicitly demonstrate that, satisfying the perfect correlation
form (\ref{5}) of the original Bell inequality for any three qubit
observables (with operator norms $\left\Vert \cdot \right\Vert \leq 1)$, the
separable noisy singlet $\eta _{\psi ^{(-)}}(\beta )=\beta \frac{I_{\mathbb{C%
}^{2}\otimes \mathbb{C}^{2}}}{4}+(1-\beta )|\psi ^{(-)}\rangle \langle \psi
^{(-)}|,$ $\beta \in \lbrack \frac{2}{3},1),$ does not exhibit Bell's
perfect correlations (see footnote 8).

Recall that, in an orthonormal basis $\{e_{k},$ $k=1,2\}$ in $\mathbb{C}^{2}$%
, a generic qubit observable has the form $W_{\alpha ,n}=$ $\alpha I_{%
\mathbb{C}^{2}}+n_{x}\sigma _{x}+$ $n_{y}\sigma _{y}+$ $n_{z}\sigma _{z},$
where: (i) $\alpha \ $is any real number; (ii) $\sigma _{x}:=|e_{1}\rangle
\langle e_{2}|$ $+$ $|e_{2}\rangle \langle e_{1}|$, $\sigma
_{y}:=i(|e_{2}\rangle \langle e_{1}|$ $-$ $|e_{1}\rangle \langle e_{2}|),$ $%
\sigma _{z}:=|e_{1}\rangle \langle e_{1}|$ $-$ $|e_{2}\rangle \langle e_{2}|$
are self-adjoint operators, represented in a basis $\{e_{k}\}$ by the Pauli
matrices; (iii) $n=(n_{x},n_{y},n_{z})$ is a vector in $\mathbb{R}^{3}$. The
eigenvalues of $W_{\alpha ,n}$ are equal to $\alpha \pm |n|,$ where $|n|:=%
\sqrt{n_{x}^{2}+n_{y}^{2}+n_{z}^{2}}.$

For the noisy singlet $\eta _{\psi ^{(-)}}(\beta )$, the correlation
function 
\begin{equation}
\mathrm{tr}[\eta _{\psi ^{(-)}}(\beta )(W_{\alpha ,n}\otimes W_{\alpha
,n})]=\alpha ^{2}-\left\vert n\right\vert ^{2}(1-\beta )  \label{18}
\end{equation}%
is negative\footnote{%
Negativity of the correlation function (\ref{18}) rules out Bell's perfect
correlations.} for any qubit observable $W_{\alpha ,n}$ with $\alpha
^{2}<\left\vert n\right\vert ^{2}(1-\beta )$, in particular, for any spin
observable\footnote{%
For a spin observable, $\alpha =0,$ $\left\vert n\right\vert =1.$}.

Furthermore, in case of Alice and Bob projective measurements of the same
qubit observable $W_{\alpha ,n},$ $\left\vert n\right\vert \neq 0$, in the
noisy singlet $\eta _{\psi ^{(-)}}(\beta )$ the joint probabilities
constitute: 
\begin{eqnarray}
\mathrm{Prob}\{\lambda _{1} &=&\alpha \pm \left\vert n\right\vert ,\text{ }%
\lambda _{2}=\alpha \pm \left\vert n\right\vert \}=\frac{\beta }{4},
\label{19} \\
\mathrm{Prob}\{\lambda _{1} &=&\alpha \pm \left\vert n\right\vert ,\text{ }%
\lambda _{2}=\alpha \mp \left\vert n\right\vert \}=\frac{1}{2}-\frac{\beta }{%
4}.  \notag
\end{eqnarray}%
Hence, given an outcome, say of Bob, the conditional probability that Alice
observes a different outcome is equal to $1-\frac{\beta }{2}$ while the
conditional probability that Alice observes the same outcome is $\frac{\beta 
}{2}.$

Thus, satisfying the perfect correlation form (\ref{5}) of the original Bell
inequality under Alice and Bob projective measurements of any three qubit
observables, the separable noisy singlet $\eta _{\psi ^{(-)}}(\beta ),$ $%
\beta \in \lbrack \frac{2}{3},1),$ \emph{does} \emph{not} exhibit the
perfect correlation of outcomes whenever the same qubit observable is
measured on both "sides".

\section{Appendix}

Below, in theorem 3, we exploit the following property of probability
measures on product spaces.

Let $\pi $ be a probability distribution with outcomes in $\Lambda
_{1}\times \Lambda _{2}\times \Lambda _{2}.$ For any subsets\footnote{%
For simplicity, we do not specify here the notion of a $\sigma $-algebra of
subsets of $\Lambda .$} $B_{1}\subseteq \Lambda _{1},$ $B^{\prime }\subseteq
\Lambda _{2}\times \Lambda _{2},$ the relation $\pi (\Lambda _{1}\times
B^{\prime })=0$ implies $\pi (B_{1}\times B^{\prime })=0.$ Hence, for any $%
B_{1}\subseteq \Lambda _{1},$ the probability distribution $\pi (B_{1}\times
\cdot )$ of outcomes in $\Lambda _{2}\times \Lambda _{2}$ is absolutely
continuous\footnote{%
On this notion, see, for example, [10].} with respect to the marginal
probability distribution $\pi (\Lambda _{1}\times \cdot )$.

The latter implies that, for any subsets $B_{1}\subseteq \Lambda _{1},\
B_{2},$ $\widetilde{B}_{2}\subseteq \Lambda _{2},$ the probability
distribution $\pi (B_{1}\times B_{2}\times \widetilde{B}_{2})$ admits the
representation 
\begin{equation}
\pi (B_{1}\times B_{2}\times \widetilde{B}_{2})=\int_{B_{2}\times \widetilde{%
B}_{2}}\nu (B_{1}|\lambda _{2},\widetilde{\lambda }_{2})\pi (\Lambda
_{1}\times d\lambda _{2}\times d\widetilde{\lambda }_{2}),  \tag{A1}
\end{equation}
where: (i) for $\pi $-almost all $\lambda _{2},$ $\widetilde{\lambda }%
_{2}\in \Lambda _{2},$ the mapping $\nu (\cdot |\lambda _{2},\widetilde{%
\lambda }_{2})$ is a probability distribution of outcomes in $\Lambda _{1}$;
(ii) for any subset $B_{1}\subseteq \Lambda _{1},$ the real-valued function $%
\nu (B_{1}|\cdot ,\cdot ):\Lambda _{2}\times \Lambda _{2}\rightarrow \lbrack
0,1]$ is measurable.

\begin{theorem}
A DSO state $\rho $ on $\mathcal{H}_{1}\otimes \mathcal{H}_{2}$ satisfies
the extended CHSH inequality (\ref{6}) under any generalized quantum
measurements of Alice and Bob, with real-valued outcomes $\left| \lambda
_{1}\right| \leq 1,$ $\left| \lambda _{2}\right| \leq 1$ of an arbitrary
spectral type.
\end{theorem}

\begin{proof}
Let a DSO state $\rho $ have a density source operator $T_{%
\blacktriangleright }$ on $\mathcal{H}_{1}\otimes \mathcal{H}_{2}\otimes 
\mathcal{H}_{2}$. Due to the definition (see section 2) of a DSO$\
T_{\blacktriangleright }$, we have:%
\begin{eqnarray}
\mathrm{tr}[\rho (M_{1}^{(a)}(d\lambda _{1})\otimes M_{2}^{(b_{1})}(d\lambda
_{2}))] &=&\mathrm{tr}[T_{\blacktriangleright }(M_{1}^{(a)}(d\lambda
_{1})\otimes M_{2}^{(b_{1})}(d\lambda _{2})\otimes I_{\mathcal{H}_{2}})], 
\TCItag{A2} \\
\mathrm{tr}[\rho (M_{1}^{(a)}(d\lambda _{1})\otimes M_{2}^{(b_{2})}(d\lambda
_{2}))] &=&\mathrm{tr}[T_{\blacktriangleright }(M_{1}^{(a)}(d\lambda
_{1})\otimes I_{\mathcal{H}_{2}}\otimes M_{2}^{(b_{2})}(d\lambda _{2}))]. 
\notag
\end{eqnarray}%
Due to the normalization of a POV measure, the probability distributions,
standing in the right-hand sides of (A2), constitute the marginals of the
probability distribution $\mathrm{tr}[T_{\blacktriangleright
}(M_{1}^{(a)}(d\lambda _{1})\otimes M_{2}^{(b_{1})}(d\lambda _{2})\otimes
M_{2}^{(b_{2})}(d\lambda _{2}))].$ The latter and the representation (\ref{1}%
) imply: 
\begin{eqnarray}
&&\text{ }\gamma _{11}\langle \lambda _{1}\lambda _{2}\rangle _{\rho
}^{(a_{1},b_{1})}+\gamma _{12}\langle \lambda _{1}\lambda _{2}\rangle _{\rho
}^{(a_{1},b_{2})}+\gamma _{21}\langle \lambda _{1}\lambda _{2}\rangle _{\rho
}^{(a_{2},b_{1})}+\gamma _{22}\langle \lambda _{1}\lambda _{2}\rangle _{\rho
}^{(a_{2},b_{2})}  \TCItag{A3} \\
&=&\int_{\Lambda _{1}\times \Lambda _{2}\times \Lambda _{2}}\text{ }(\gamma
_{11}\lambda _{1}\lambda _{2}+\gamma _{12}\lambda _{1}\widetilde{\lambda }%
_{2})\mathrm{tr}[T_{\blacktriangleright }(M_{1}^{(a_{1})}(d\lambda
_{1})\otimes M_{2}^{(b_{1})}(d\lambda _{2})\otimes M_{2}^{(b_{2})}(d%
\widetilde{\lambda }_{2}))]  \notag \\
&&+\int_{\Lambda _{1}\times \Lambda _{2}\times \Lambda _{2}}\text{ }(\gamma
_{21}\widetilde{\lambda }_{1}\lambda _{2}+\gamma _{22}\widetilde{\lambda }%
_{1}\widetilde{\lambda }_{2})\mathrm{tr}[T_{\blacktriangleright
}(M_{1}^{(a_{2})}(d\widetilde{\lambda }_{1})\otimes M_{2}^{(b_{1})}(d\lambda
_{2})\otimes M_{2}^{(b_{2})}(d\widetilde{\lambda }_{2}))],  \notag
\end{eqnarray}%
where $\Lambda _{1},\Lambda _{2}\subseteq \lbrack -1,1].$ Due to (A1), we
have: 
\begin{eqnarray}
&&\mathrm{tr}[T_{\blacktriangleright }(M_{1}^{(a_{1})}(d\lambda _{1})\otimes
M_{2}^{(b_{1})}(d\lambda _{2})\otimes M_{2}^{(b_{2})}(d\widetilde{\lambda }%
_{2}))]  \TCItag{A4} \\
&=&\nu _{a_{1},b_{1},b_{2}}(d\lambda _{1}|\lambda _{2},\widetilde{\lambda }%
_{2})\text{ }\mathrm{tr}[T_{\blacktriangleright }(I_{\mathcal{H}_{1}}\otimes
M_{2}^{(b_{1})}(d\lambda _{2})\otimes M_{2}^{(b_{2})}(d\widetilde{\lambda }%
_{2}))],  \notag \\
&&  \notag \\
&&\mathrm{tr}[T_{\blacktriangleright }(M_{1}^{(a_{2})}(d\widetilde{\lambda }%
_{1})\otimes M_{2}^{(b_{1})}(d\lambda _{2})\otimes M_{2}^{(b_{2})}(d%
\widetilde{\lambda }_{2}))]  \notag \\
&=&\nu _{a_{2},b_{1},b_{2}}(d\widetilde{\lambda }_{1}|\lambda _{2},%
\widetilde{\lambda }_{2})\text{ }\mathrm{tr}[T_{\blacktriangleright }(I_{%
\mathcal{H}_{1}}\otimes M_{2}^{(b_{1})}(d\lambda _{2})\otimes
M_{2}^{(b_{2})}(d\widetilde{\lambda }_{2}))],  \notag
\end{eqnarray}%
where $\nu _{a_{1},b_{1},b_{2}}(\cdot |\lambda _{2},\widetilde{\lambda }%
_{2}) $ and $\nu _{a_{2},b_{1},b_{2}}(\cdot |\lambda _{2},\widetilde{\lambda 
}_{2}) $ are probability distributions of outcomes in $\Lambda _{1}$.
Introducing the probability distribution 
\begin{eqnarray}
&&\mu (d\lambda _{1}\times d\widetilde{\lambda }_{1}\times d\lambda
_{2}\times d\widetilde{\lambda }_{2})  \TCItag{A5} \\
&=&\nu _{a_{1},b_{1},b_{2}}(d\lambda _{1}|\lambda _{2},\widetilde{\lambda }%
_{2})\nu _{a_{2},b_{1},b_{2}}(d\widetilde{\lambda }_{1}|\lambda _{2},%
\widetilde{\lambda }_{2})\mathrm{tr}[T_{\blacktriangleright }(I_{\mathcal{H}%
_{1}}\otimes M_{2}^{(b_{1})}(d\lambda _{2})\otimes M_{2}^{(b_{2})}(d%
\widetilde{\lambda }_{2}))],  \notag
\end{eqnarray}%
we rewrite (A3) in the form: 
\begin{eqnarray}
&&\gamma _{11}\langle \lambda _{1}\lambda _{2}\rangle _{\rho
}^{(a_{1},b_{1})}+\gamma _{12}\langle \lambda _{1}\lambda _{2}\rangle _{\rho
}^{(a_{1},b_{2})}+\gamma _{21}\langle \lambda _{1}\lambda _{2}\rangle _{\rho
}^{(a_{2},b_{1})}+\gamma _{22}\langle \lambda _{1}\lambda _{2}\rangle _{\rho
}^{(a_{2},b_{2})}  \TCItag{A6} \\
&=&\int_{\Lambda _{1}\times \Lambda _{1}\times \Lambda _{2}\times \Lambda
_{2}}(\gamma _{11}\lambda _{1}\lambda _{2}+\gamma _{12}\lambda _{1}%
\widetilde{\lambda }_{2}+\gamma _{21}\widetilde{\lambda }_{1}\lambda
_{2}+\gamma _{22}\widetilde{\lambda }_{1}\widetilde{\lambda }_{2})\mu
(d\lambda _{1}\times d\widetilde{\lambda }_{1}\times d\lambda _{2}\times d%
\widetilde{\lambda }_{2}).  \notag
\end{eqnarray}%
Taking into account that, for any real numbers $\left\vert \lambda
_{1}\right\vert \leq 1,$ $\left\vert \lambda _{2}\right\vert \leq 1,$ the
inequality 
\begin{equation}
\left\vert \gamma _{11}\lambda _{1}\lambda _{2}+\gamma _{12}\lambda _{1}%
\widetilde{\lambda }_{2}+\gamma _{21}\widetilde{\lambda }_{1}\lambda
_{2}+\gamma _{22}\widetilde{\lambda }_{1}\widetilde{\lambda }_{2}\right\vert
\leq 2\max_{i,k}\left\vert \gamma _{ik}\right\vert  \tag{A7}
\end{equation}%
holds with any real coefficients $\gamma _{ik}$, satisfying either of the
relations in (\ref{7}), we derive: 
\begin{eqnarray}
&&\left\vert \gamma _{11}\langle \lambda _{1}\lambda _{2}\rangle _{\rho
}^{(a_{1},b_{1})}+\gamma _{12}\langle \lambda _{1}\lambda _{2}\rangle _{\rho
}^{(a_{1},b_{2})}+\gamma _{21}\langle \lambda _{1}\lambda _{2}\rangle _{\rho
}^{(a_{2},b_{1})}+\gamma _{22}\langle \lambda _{1}\lambda _{2}\rangle _{\rho
}^{(a_{2},b_{2})}\right\vert  \TCItag{A8} \\
&\leq &\int_{\Lambda _{1}\times \Lambda _{1}\times \Lambda _{2}\times
\Lambda _{2}}\left\vert \gamma _{11}\lambda _{1}\lambda _{2}+\gamma
_{12}\lambda _{1}\widetilde{\lambda }_{2}+\gamma _{21}\widetilde{\lambda }%
_{1}\lambda _{2}+\gamma _{22}\widetilde{\lambda }_{1}\widetilde{\lambda }%
_{2}\right\vert \mu (d\lambda _{1}\times d\widetilde{\lambda }_{1}\times
d\lambda _{2}\times d\widetilde{\lambda }_{2})  \notag \\
&\leq &2\max_{i,k}\left\vert \gamma _{ik}\right\vert .  \notag
\end{eqnarray}%
The validity of (\ref{6}) for a DSO state $\rho $ that has a density
source-operator $T_{\blacktriangleleft }$ is proved quite similarly.
\end{proof}

\end{document}